\documentclass[aps,prl,singlecolumn,superscriptaddress,nofootinbib,floatfix]{revtex4-2}
\usepackage{amsmath}
\usepackage{graphicx}
\usepackage{dcolumn}
\usepackage{bm}
\usepackage{amssymb}
\usepackage{dsfont}
\usepackage{url}
\usepackage{caption,subcaption}
\usepackage[colorlinks=true,allcolors=blue]{hyperref}

\begin{document}

\title{Novel Relaxation Time Approximation to the Relativistic Boltzmann Equation}
\author{Gabriel S. Rocha}
\email{gabrielsr@id.uff.br}
\affiliation{Instituto de F\'{\i}sica, Universidade Federal Fluminense, Niter\'{o}i, Rio
de Janeiro, Brazil}
\author{Gabriel S. Denicol}
\email{gsdenicol@id.uff.br}
\affiliation{Instituto de F\'{\i}sica, Universidade Federal Fluminense, Niter\'{o}i, Rio
de Janeiro, Brazil}
\author{Jorge Noronha} 
\email{jn0508@illinois.edu}
\affiliation{Illinois Center for Advanced Studies of the Universe\\ Department of Physics, 
University of Illinois at Urbana-Champaign, Urbana, IL 61801, USA}

\begin{abstract}
We show that the widely used relaxation time approximation to the relativistic Boltzmann equation contains basic flaws, being incompatible with microscopic and macroscopic conservation laws. We propose a new approximation that fixes such fundamental issues and maintains the basic properties of the linearized Boltzmann collision operator. We show how this correction affects transport coefficients, such as the bulk viscosity and particle diffusion.
\end{abstract}

\maketitle
\noindent
\textit{Introduction:} In 1974, Anderson and Witting (AW) proposed the relaxation time approximation (RTA) to the relativistic Boltzmann equation \cite{AW}, following the development already made in the non-relativistic case by Bhatnagar, Gross and Krook \cite{bgk} and Welander \cite{welander}. This approximation is used in several fields of physics and has been recently employed to study the hydrodynamization of the matter produced in ultrarelativistic heavy-ion collisions \cite{Florkowski:2017olj,Denicol:2014xca,Denicol:2014tha,Heller:2015dha,Romatschke:2017vte,Strickland:2017kux,Behtash:2017wqg,Denicol:2017lxn,Blaizot:2017ucy,Heller:2018qvh,Almaalol:2018ynx,Denicol:2018pak,Behtash:2018moe,Behtash:2019txb,Strickland:2018ayk,Strickland:2019hff,Kurkela:2019set,Denicol:2019lio}. In particular, the divergence of the hydrodynamic gradient expansion was first shown in relativistic kinetic theory using the relaxation time approximation \cite{Denicol:2016bjh,Heller:2016rtz}, in agreement with previous calculations at strong coupling \cite{Heller:2013fn,Buchel:2016cbj}. For this reason, the relaxation time approximation has become instrumental in studying the microscopic foundations and the domain of applicability of relativistic hydrodynamics. 

In this letter we show that, despite its wide use, the approximation proposed in \cite{AW} contains basic flaws. It is incompatible with microscopic and macroscopic conservation laws, which leads to several problems when modeling relativistic gases using energy dependent relaxation times or general matching conditions. We propose a new relaxation time approximation that fixes such fundamental issues and maintains the basic properties of the linearized Boltzmann collision operator. We then demonstrate how such new formulation modifies well-known results for the transport coefficients present in the first-order Chapman-Enskog expansion \cite{chapman-cowling}, such as the shear and bulk viscosities, as well as the particle/heat diffusion coefficient. Our new approach provides a simple way to consistently investigate different out of equilibrium definitions of the hydrodynamics fields in relativistic kinetic theory, which are not possible using the standard AW formulation. 

In this work we use the mostly minus convention for the
metric, $g^{\mu\nu}=\textrm{diag}(1,-1,-1,-1)$, and natural units, i.e. $\hbar=c=k_{B}=1$. Throughout the text, the spacetime dependence of some
functions is omitted and the momentum dependence is denoted by a sub-index,
so that $f(x,p)=f_{\mathbf{p}}$.\\

\noindent
\textit{Boltzmann equation:} The relativistic Boltzmann equation is an
integro-differential equation for the single-particle momentum distribution
function $f_{\mathbf{p}}$. For a one-component gas composed of \textit{classical}
particles\footnote{For simplicity, here we only consider classical statistics. The inclusion of quantum statistics effects is straightforward.} that only undergo elastic scattering, it reads 
\begin{equation}
k^{\mu }\partial _{\mu }f_{\mathbf{k}}=\int dP\ dP^{\prime }\ dK^{\prime }%
\,\mathcal{W}_{\mathbf{kk}^{\prime }\leftrightarrow \mathbf{pp}^{\prime }}(f_{%
\mathbf{p}}f_{\mathbf{p}^{\prime }}-f_{\mathbf{k}}f_{\mathbf{k}^{\prime
}})\equiv C\left[ f\right] ,  \label{eq:BE}
\end{equation}%
where the right-hand side displays the so-called collision term or collision
integral, which can be readily seen as the most nontrivial part of the Boltzmann equation 
\cite{cercignani:90mathematical}. Above, $\mathcal{W}_{\mathbf{kk}^{\prime
}\leftrightarrow \mathbf{pp}^{\prime }}$ is the transition rate, $k^{\mu
}=\left( k^{0},\mathbf{k}\right) $ is the 4-momentum, and we introduced the Lorentz
invariant integration measure $dP\equiv d^{3}p/\left[ p^{0}\left( 2\pi
\right) ^{3}\right]$.

When considering systems that are close to local equilibrium, such as
relativistic fluids, a common and useful approximation is to linearize the
collision integral around a local equilibrium state, $f_{0\mathbf{k}}=\exp
\left( -\beta u_{\mu }k^{\mu }+\alpha \right) $, with $\beta $ being the
inverse temperature, $\alpha $ the thermal potential, and $u_{\mu }$ a
normalized 4-velocity (i.e. $u_\mu u^\mu = 1$). In this case, one expresses $f_{\mathbf{k}}$ in terms
of an equilibrium contribution and a nonequilibrium correction, $\phi _{%
\mathbf{k}}$, as 
\begin{equation}
f_{\mathbf{k}}\equiv f_{0\mathbf{k}}\left( 1+\phi _{\mathbf{k}}\right) ,
\end{equation}%
and, neglecting terms that are quadratic in $\phi _{\mathbf{k}}$, one obtains
the following approximate form for the Boltzmann equation%
\begin{equation}
k^{\mu }\partial _{\mu }f_{\mathbf{k}}=\hat{L}\phi _{\mathbf{k}}.
\end{equation}%
Here, we introduced the \textit{linearized} collision operator, $\hat{L}$,
which is defined in the following way
\begin{equation}
\hat{L}\phi _{\mathbf{k}}\equiv \int dP\ dP^{\prime }\ dK^{\prime }\,\mathcal{W%
}_{\mathbf{kk}^{\prime }\leftrightarrow \mathbf{pp}^{\prime }}f_{0\mathbf{k}%
}f_{0\mathbf{k}^{\prime }}(\phi _{\mathbf{p}}+\phi _{\mathbf{p}^{\prime
}}-\phi _{\mathbf{k}}-\phi _{\mathbf{k}^{\prime }}).
\end{equation}%
We remark that the inverse temperature, thermal potential, and $4$--velocity are not intrinsically part of the Boltzmann equation, since they were introduced through this linearization procedure. These variables are commonly defined using the so-called matching conditions \cite{cercignani:90mathematical,Israel:1979wp,degroot}.

If interpreted as a linear operator in Hilbert space, $\hat{L}$ can be shown
to be a negative semidefinite operator \cite{cercignani:90mathematical}. These properties
imply that $\hat{L}$ has only non-positive eigenvalues, the
absolute values of which may be interpreted as the reciprocal of the
microscopic relaxation times of non-equilibrium perturbations \cite%
{reichl:99,cercignani:02relativistic}. A fundamental property of the
linearized Boltzmann collision operator, valid for any type of interactions, is that its only eigenfunctions with zero eigenvalues are the quantities
that are conserved in microscopic collisions such as energy, momentum, and, in our particular case, particle number. That is, 
\begin{equation}
\hat{L}1=0\left. {}\right. \mathrm{and}\left. {}\right. \hat{L}k^{\mu }=0.
\end{equation}%
Furthermore, fundamental properties of the transition probability rate $%
\mathcal{W}_{\mathbf{kk}^{\prime }\leftrightarrow \mathbf{pp}^{\prime }}$
(such as time reversal symmetry) guarantee that $\hat{L}$ is self-adjoint%
\begin{equation}
\int dK\ \psi _{\mathbf{k}}\hat{L}\phi _{\mathbf{k}}=\int dK\ \phi _{\mathbf{%
k}}\hat{L}\psi _{\mathbf{k}},
\end{equation}%
and, thus, this integral also vanishes if $\psi _{\mathbf{k}}$ corresponds
to a quantity that is conserved in collisions, i.e., $\int dK
\hat{L}\phi _{%
\mathbf{k}}=0$ and $\int dK\ k^{\mu }\hat{L}\phi _{\mathbf{k}}=0$.

The properties above are of the utmost importance in the derivation of
macroscopic conservation laws from the linearized Boltzmann equation. Continuity equations that describe conservation of particle number and
energy-momentum are obtained by multiplying both sides of the linearized
Boltzmann equation by $1$ and $k^{\mu }$ and integrating in momentum, thus
obtaining%
\begin{equation}
\partial_{\mu }\left\langle k^{\mu }\right\rangle =0,\left. {}\right.
\partial_{\mu }\left\langle k^{\mu }k^{\nu }\right\rangle =0, \label{cont}
\end{equation}%
where we made use of the following notation 
\begin{equation}
\left\langle \cdots \right\rangle \equiv \int dK\left. {}\right. \left(
\cdots \right) f_{\mathbf{k}}.
\end{equation}
From (\ref{cont}), one identifies $\left\langle
k^{\mu }\right\rangle $ as the conserved particle 4-current and $\left\langle k^{\mu }k^{\nu }\right\rangle $ as the energy-momentum tensor.\\

\noindent
\textit{Relaxation time approximation:} The relaxation time approximation
corresponds to a simplification of the linearized collision operator, $\hat{L}$. The
approximation proposed by Anderson and Witting amounts to \cite{AW} 
\begin{equation}
\hat{L}\phi _{\mathbf{k}}\approx \hat{L}_{\mathrm{RTA}}\phi _{\mathbf{k}%
}\equiv -\frac{E_{\mathbf{k}}}{\tau _{R}}f_{0\mathbf{k}}\phi _{\mathbf{k}},
\label{eq:approx-RTA-trad}
\end{equation}%
where $E_{\mathbf{k}}\equiv u_{\mu }k^{\mu }$ is the energy of the particle in the local rest frame and $\tau _{R}=\tau _{R}\left( E_{\mathbf{k}}\right) $ is the relaxation
time, which is interpreted as a phenomenological time scale within which the system reaches equilibrium. This model is already an improvement over
the approximation proposed by Marle \cite{Marle}, where it was imposed  $\hat{%
L}_{\mathrm{RTA}}\phi _{\mathbf{k}}\approx -\left( m/\tau _{R}\right) f_{0%
\mathbf{k}}\phi _{\mathbf{k}}$, with $m$ being the particle on-shell mass. Marle's model is obviously ill-defined in
the ultrarelativistic limit while the approximation proposed by Anderson
and Witting was able to recover qualitatively the results obtained using
Grad's method of moments in the ultrarelativistic limit \cite{AW}.
Furthermore, the AW formulation also makes it possible to find analytical and semi-analytical
solutions for $f_{\mathbf{k}}$ \cite{Denicol:2014xca,Denicol:2014tha,Noronha:2015jia,Hatta:2015kia,Florkowski:2013lya,Bazow:2015dha,Bazow:2016oky}. We
note that, in Hilbert space, the AW prescription corresponds to setting 
$\hat{L}_{\mathrm{RTA}}\sim -\mathds{1}$, with $\mathds{1}$ being
the identity operator.

However, this approximation of the collision operator contains basic flaws
that must be addressed -- this being the main motivation of this letter. The
main problem is that this formulation is not consistent with the fundamental properties of the collision operator discussed above.
As a matter of fact, instead of \eqref{cont}, one obtains the following equations of motion for the conserved currents using the AW approximation,
\begin{eqnarray}
\partial _{\mu }\left\langle k^{\mu }\right\rangle &=&-\int dK\ \frac{E_{%
\mathbf{k}}}{\tau _{R}}f_{0\mathbf{k}}\phi _{\mathbf{k}}, \label{eq:csv-RTA-trad0} \\
\partial _{\mu }\left\langle k^{\mu }k^{\nu }\right\rangle &=&-\int dK\
k^{\mu }\frac{E_{\mathbf{k}}}{\tau _{R}}f_{0\mathbf{k}}\phi _{\mathbf{k}}. \label{eq:csv-RTA-trad1}
\end{eqnarray}%
In many applications, the right-hand side of Eqs.~\eqref{eq:csv-RTA-trad0} and \eqref{eq:csv-RTA-trad1} is set to zero by imposing the Landau matching conditions 
\cite{landau},
\begin{equation}
\int dK\ E_{\mathbf{k}}f_{0\mathbf{k}}\phi _{\mathbf{k}}=0,\int dK\ k^{\mu
}E_{\mathbf{k}}f_{0\mathbf{k}}\phi _{\mathbf{k}}=0, \label{matchingL}
\end{equation}%
which are used to define the temperature, thermal potential, and 4-velocity
introduced in the relaxation time approximation (\ref{eq:approx-RTA-trad}). Nevertheless, even this
procedure is not general as it only guarantees the validity of the
conservation laws if one imposes that the relaxation time, $\tau_R$, has no momentum dependence. Therefore, in order to circumvent such a fundamental problem, another procedure to approximate the linearized collision term becomes necessary.

We propose the following modification of the
relaxation time approximation to resolve this problem%
\begin{equation}
\hat{L}_{\mathrm{RTA}}\sim -\mathds{1}\longrightarrow \hat{L}_{\mathrm{RTA}%
}\sim -\mathds{1}+\sum_{n=1}^{5}|\lambda_n\rangle \langle \lambda_n|.  \label{newRTA}
\end{equation}%
The expressions above are written in Hilbert space, with $|\lambda_n\rangle $ being the 5 degenerate \textit{orthonormal}
eigenvectors of $\hat{L}$ that have a vanishing eigenvalue, i.e., $\hat{L}%
|\lambda_n\rangle =0$. This is nothing but the usual relaxation time approximation
combined with counterterms in such a way that it becomes a projector onto
the subspace orthogonal to $|\lambda_n\rangle $. Naturally, with this modification,
one guarantees that $\hat{L}_{\mathrm{RTA}}|\lambda_n\rangle =0$ independently of
any matching condition or energy dependence of the relaxation time, as it occurs for the linearized Boltzmann equation. The solution proposed above
is well-known in the non-relativistic limit \cite%
{reichl:99,cercignani:90mathematical}, but it has never been applied in the
relativistic regime.

Our goal is now to write \eqref{newRTA} in a less abstract form. In order to do so, we first have to re-express the microscopic conserved quantities, $1$ and $k^{\mu }$, as an orthogonal
basis:
\begin{equation}
P_{0}^{\left( 0\right) }\equiv 1,\left. {}\right. P_{1}^{\left( 0\right)
}\equiv 1-\frac{\left\langle E_{\mathbf{k}}/\tau _{R} \right\rangle _{0}}{\left\langle E_{\mathbf{k}}^2/\tau _{R} \right\rangle _{0}} E_{\mathbf{k}},\left. {}\right.
k^{\left\langle \mu \right\rangle }\equiv \Delta ^{\mu \nu }k_{\nu }\text{,}
\end{equation}%
where $\Delta^{\mu\nu} \equiv g^{\mu\nu} - u^{\mu}u^{\nu}$ is a projection operator onto the 3-space orthogonal to $u^{\mu}$, and we defined the momentum
integrals relative to the local equilibrium distribution function $f_{0\mathbf{k}}$ as follows
\begin{equation}
\left\langle \cdots \right\rangle _{0}\equiv \int dK f_{0\mathbf{k}}\left( \cdots \right) .
\end{equation}%
These basis elements are constructed to satisfy the following orthogonality relations,%
\begin{equation}
\left\langle \left( E_{\mathbf{k}}/\tau _{R}\right) P_{0}^{\left( 0\right) }P_{1}^{\left( 0\right) }\right\rangle
_{0}=0,\left. {}\right. \left\langle \left( E_{\mathbf{k}}/\tau _{R}\right) P_{0}^{\left( 0\right) }k^{\left\langle
\mu \right\rangle }\right\rangle _{0}=0,\left. {}\right. \left\langle
\left( E_{\mathbf{k}}/\tau _{R}\right) P_{1}^{\left( 0\right) }k^{\left\langle \mu \right\rangle }\right\rangle
_{0}=0.
\end{equation}%
We then rewrite the relaxation time approximation in Eq.~\eqref{newRTA} using this basis, which gives the following expression,%
\begin{equation}
\hat{L}_{\mathrm{RTA}}\phi _{\mathbf{k}}=-\frac{E_{\mathbf{k}}}{\tau _{R}}%
f_{0\mathbf{k}}\left[ \phi _{\mathbf{k}}-\frac{\left\langle \left( E_{\mathbf{k}}/\tau _{R}\right) \phi _{\mathbf{k}%
}\right\rangle _{0}}{\left\langle E_{\mathbf{k}}/\tau _{R}\right\rangle _{0}}-P_{1}\frac{%
\left\langle \left( E_{\mathbf{k}}/\tau _{R}\right) P_{1}^{\left( 0\right) }\phi _{\mathbf{k}}\right\rangle _{0}}{%
\left\langle \left( E_{\mathbf{k}}/\tau _{R}\right) P_{1}^{\left( 0\right) }P_{1}^{\left( 0\right) }\right\rangle
_{0}}-k^{\left\langle \mu \right\rangle }\frac{\left\langle \left( E_{\mathbf{k}}/\tau _{R}\right) k_{\left\langle
\mu \right\rangle }\phi_{\mathbf{k}}\right\rangle _{0}}{(1/3)
\left\langle \left( E_{\mathbf{k}}/\tau _{R}\right) k_{\left\langle \nu \right\rangle }k^{\left\langle \nu
\right\rangle }\right\rangle _{0}}\right] .
\end{equation}%
This is the main result of this letter. In this form, $\hat{L}_{\mathrm{RTA}}$ is an integral operator and, in this
sense, becomes harder to invert when compared to the usual Anderson-Witting approximation. We note that a reasoning similar to what we have presented above has also
been employed to render the relaxation time approximation compatible with
the Eckart frame in Refs.\ \cite{Pennisi:18,Carrisi:19}.\\

\noindent
\textit{Chapman-Enskog expansion:} A natural step after the derivation of this new approximation for the linearized collision operator is to analyze the
effects of the energy dependence of the relaxation time and the choice of
matching conditions in the hydrodynamic regime. This will be done in the
following by employing the traditional Chapman-Enskog series -- a derivative expansion of solutions of the Boltzmann equation \cite{chapman-cowling,cercignani:02relativistic,cercignani:90mathematical}. The zeroth-order solution of such an expansion leads to the local equilibrium solution itself and, thus, to $\phi _{\mathbf{k}}=0$. On the other hand, the first-order Chapman-Enskog approximation for $\phi _{\mathbf{k}}$ is nontrivial and is obtained from the following equation,
\begin{equation}
f_{0\mathbf{k}}\left( A_{\mathbf{k}}\theta +B_{\mathbf{k}}k_{\mu}\nabla ^{\mu }\alpha -\beta k_{\mu}k_{\nu}\sigma ^{\mu \nu }\right) 
=\hat{L}\phi_{\mathbf{k}} \approx \hat{L}_{\mathrm{RTA}}\phi _{\mathbf{k}} , \label{maineq}
\end{equation}%
where $\theta \equiv \partial_{\mu }u^{\mu }$ is the expansion rate, $\nabla^{\mu } \equiv \Delta^{\mu\nu}\partial_{\nu }$, and $\sigma^{\mu\nu} \equiv \Delta^{\mu\nu}_{\alpha\beta} \partial_{\alpha }u_{\beta }$ is the shear tensor, with $\Delta^{\mu\nu}_{\alpha\beta} \equiv (\Delta^{\mu}_{\alpha}\Delta^{\nu}_{\beta} + \Delta^{\mu}_{\beta}\Delta^{\nu}_{\alpha})/2 - (1/3)\Delta^{\mu\nu}\Delta_{\alpha\beta}$ being a doubly symmetric and traceless projection operator. For the sake of convenience, we further introduced the following scalar functions of $\alpha ,$ $\beta $, $m$
and $E_{\mathbf{k}}$,
\begin{eqnarray*}
A_{\mathbf{k}} &=&\frac{\left( I_{20}+I_{21}\right) I_{20}-I_{10}I_{30}}{%
I_{10}I_{30}-I_{20}^{2}}E_{\mathbf{k}}-\frac{I_{10}I_{21}}{%
I_{10}I_{30}-I_{20}^{2}}E_{\mathbf{k}}^{2}-\frac{\beta }{3}\left( m^{2}-E_{%
\mathbf{k}}^{2}\right) . \\
B_{\mathbf{k}} &=&1-\frac{I_{10} E_{\mathbf{k}}}{I_{20}+I_{21}},\left. {}\right.
I_{nq}=\frac{1}{\left( 2q+1\right) !!} \left\langle E_{\mathbf{k}%
}^{n-2q}\left( m^{2}-E_{\mathbf{k}}^{2}\right)^{q} \right\rangle_{0}.
\end{eqnarray*}%
Note that the first-order Chapman-Enskog approximation leads to an integral equation for $\phi _{\mathbf{k}}$, which must be inverted. We further note that the thermodynamic functions $I_{nq}$ were previously
defined and used in Refs.~\cite{Israel:1979wp,Denicol:2012cn,Denicol:2012es} and they appear often in the derivation of fluid dynamics from the Boltzmann equation. In fact, $I_{10}\equiv n$ is the particle number density, $I_{20}\equiv \varepsilon$ is the energy density, and $I_{21}\equiv P$ is the thermodynamic pressure. 

Since $\hat{L}_{\mathrm{RTA}}$ is a linear operator, we may write the solution for $\phi _{\mathbf{k}}$ in the following form,
\begin{equation}
\phi _{\mathbf{k}}= \mathcal{S}_{\mathbf{k}}^{\left( 0 \right) }\theta +\mathcal{S}_{\mathbf{k}}^{\left( 1 \right) }k_{\mu}\nabla ^{\mu }\alpha + \mathcal{S}_{\mathbf{k}}^{\left( 2 \right) } k_{\mu}k_{\nu}\sigma ^{\mu \nu }.
\label{anothereq}
\end{equation}%
All that remains is to determine the momentum dependence of the coefficients $\mathcal{S}_{\mathbf{k}}^{\left( \ell \right) }$. We perform this task by expanding the coefficients in terms of a complete basis  described in Refs.~\cite{degroot,cercignani:02relativistic,Denicol:2012cn},  
\begin{equation}
\mathcal{S}_{\mathbf{k}}^{\left( \ell \right) }=\sum_{n=0}^{\infty }s_{n}^{\left( \ell \right)}P_{n}^{\left( \ell \right) } ,
\end{equation}
where $P_{n}^{\left( \ell \right) }$ are orthogonal polynomials that
satisfy $\left\langle \left( E_{\mathbf{k}}/\tau _{R}\right) \left( \Delta _{\alpha \beta }k^{\alpha }k^{\beta
}\right) ^{\ell }P_{n}^{\left( \ell \right) }P_{m}^{\left( \ell \right)
}\right\rangle _{0}\sim \delta _{nm}$ and can be constructed following the
Gram-Schmidt orthogonalization process as described in Refs.~\cite{cercignani:02relativistic,Denicol:2012cn}. We note that the terms associated with the coefficients $s_{0}^{\left( 0 \right)}$, $s_{1}^{\left( 0 \right)}$, and $s_{0}^{\left( 1 \right)}$ are homogeneous solutions of Eq.~(\ref{maineq}) and, thus, they cannot be obtained from an inversion procedure. Such coefficients must be determined separately, using the matching conditions imposed to define the local equilibrium state. We note that the existence of such homogeneous solutions is an essential feature of the linearized Boltzmann equation, which is preserved in our novel relaxation time approximation.

The expansion coefficients  are then obtained using the orthogonality relations satisfied by the basis elements and shall be given bellow. For the coefficients related to the scalar contribution, we find
\begin{equation}
s_{n}^{\left( 0 \right)}=-\frac{\left\langle A_{\mathbf{k}}P_{n}^{\left( 0 \right)}\right\rangle _{0}}{%
\left\langle \left( E_{\mathbf{k}}/\tau _{R}\right)  P_{n}^{\left( 0 \right)}P_{n}^{\left( 0 \right)} \right\rangle _{0}} , {} \forall \left. {}\right. n\geq 2
\label{r1}
\end{equation} 
for the coefficients related to the vector term one obtains
\begin{equation}
s_{n}^{\left( 1 \right)}=-\frac{\left\langle \Delta _{\alpha \beta }k^{\alpha }k^{\beta
}B_{\mathbf{k}} P_{n}^{\left( 1\right)}\right\rangle _{0}}{\left\langle \left( E_{\mathbf{k}}/\tau _{R}\right) \Delta _{\alpha \beta }k^{\alpha }k^{\beta
} P_{n}^{\left( 1\right)}P_{n}^{\left( 1\right)} \right\rangle_0 }
,\forall \left. {}\right. n\geq 1
\label{r2}
\end{equation}%
and, finally, for the coefficients related to the tensor term,
\begin{equation}
s_{n}^{\left( 2 \right)}=\frac{\left\langle \left( \Delta_{\alpha \beta }k^{\alpha }k^{\beta }\right)^{2} \beta P_{n}^{\left(2\right)}\right\rangle_{0}}{ \left\langle \left( E_{\mathbf{k}}/\tau _{R}\right) \left( \Delta_{\alpha \beta }k^{\alpha }k^{\beta }\right)^{2} P_{n}^{\left( 2\right)
}P_{n}^{\left( 2\right) }\right\rangle_{0}}.
\label{r3}
\end{equation}%
We note that the main difference between our approach and the traditional RTA, proposed by Anderson and Witting, is that our solutions, (\ref{r1}) and (\ref{r2}), are \textit{not} valid for all $n$ -- the homogeneous solutions have been properly subtracted in our work. Otherwise, one lacks the freedom to employ arbitrary matching conditions. Such freedom is, as mentioned above, an intrinsic property of the linearized Boltzmann equation that is fully preserved by our new formulation.

The coefficients $s_{0}^{\left( 0 \right)}$, $s_{1}^{\left( 0 \right)}$, and $s_{0}^{\left( 1 \right)}$ can only be obtained once matching conditions are provided. We now consider a general set of matching conditions
\begin{eqnarray}
\left\langle g_{\mathbf{k}}\phi_{\mathbf{k}} \right\rangle_{0}=\left\langle h_{\mathbf{k}}\phi_{\mathbf{k}} \right\rangle_{0}=0,
\left\langle q_{\mathbf{k}}k^{\left\langle \mu \right\rangle }\phi_{\mathbf{k}} \right\rangle_{0}=0, \label{matching}
\end{eqnarray}
where $g_{\mathbf{k}}(E_\mathbf{k})$ and $h_{\mathbf{k}}(E_\mathbf{k})$ are arbitrary linearly independent functions, and $q_{\mathbf{k}}(E_\mathbf{k})$ is an arbitrary function that does not have to be linearly independent from $g_{\mathbf{k}}$ and $h_{\mathbf{k}}$. These conditions lead to the following set of equations 
\begin{eqnarray}
\left\langle  g_{\mathbf{k}}  \right\rangle_{0}s_{0}^{\left( 0 \right)}+\left\langle g_{\mathbf{k}} P_{1}^{\left( 0 \right)} \right\rangle_{0}s_{1}^{\left( 0 \right)}=-\sum_{n=2}^{\infty} \left\langle  g_{\mathbf{k}}  P_{n}^{\left(0 \right)}\right\rangle_{0}s_{n}^{\left( 0 \right)} , \\
\left\langle h_{\mathbf{k}}  \right\rangle_{0}s_{0}^{\left( 0 \right)}+\left\langle h_{\mathbf{k}} P_{1}^{\left( 0 \right)} \right\rangle_{0}s_{1}^{\left( 0 \right)}=-\sum_{n=2}^{\infty} \left\langle h_{\mathbf{k}}  P_{n}^{\left(0 \right)}\right\rangle_{0}s_{n}^{\left( 0 \right)} , \\
\left\langle  q_{\mathbf{k}} \Delta_{\alpha \beta}k^{\alpha }k^{\beta}
\right\rangle_{0}s_{0}^{\left( 1 \right)} =-\sum_{n=1}^{\infty} \left\langle q_{\mathbf{k}} \Delta _{\alpha \beta }k^{\alpha }k^{\beta
} P_{n}^{\left(1 \right)} \right\rangle_{0}s_{n}^{\left( 0 \right)}.
\end{eqnarray}%
The equations above can be solved to obtain $s_{0}^{\left( 0 \right)}$, $s_{1}^{\left( 0 \right)}$, and $s_{0}^{\left( 1 \right)}$ in terms of solutions (\ref{r1}) and (\ref{r2}). Naturally, the general solution for $\phi_{\mathbf{k}}$ will depend on the choice of the functions $g_{\mathbf{k}}$, $h_{\mathbf{k}}$, and $q_{\mathbf{k}}$, i.e., they will depend on the choice of matching conditions.\\

\noindent
\textit{Transport coefficients:} The bulk viscous pressure, $\Pi$, particle diffusion 4-current, $n^{\mu}$, and shear stress tensor, $\pi^{\mu\nu}$, are obtained by replacing the first-order Chapman-Enskog solutions for $\phi_{\mathbf{k}}$ obtained above into the definitions of such dissipative currents,  
\begin{eqnarray}
\Pi = -\frac{1}{3}  \Delta_{\alpha\beta} \left\langle k^{\alpha }k^{\beta}\phi_{\mathbf{k}}  \right\rangle_{0}, 
n^\mu = \Delta^{\mu}_{\nu} \left\langle k^{\nu} \phi_{\mathbf{k}}  \right\rangle_{0}, 
\pi^{\mu\nu} = \Delta^{\mu \nu}_{\alpha \beta} \left\langle k^{\alpha}k^{\beta} \phi_{\mathbf{k}}  \right\rangle_{0}.   \label{definitions}
\end{eqnarray}
This procedure leads to relativistic Navier-Stokes theory where, $\Pi=-\zeta \theta$, $n^\mu=\kappa_n \nabla^{\mu} \alpha$, $\pi^{\mu\nu}=2\eta \sigma^{\mu\nu}$, with $\zeta$ being the bulk viscosity, $\kappa_n$ the particle diffusion coefficient, and $\eta$ the shear viscosity. One then obtains the following microscopic expressions for these transport coefficients,
\begin{eqnarray}
\zeta &=& \frac{1}{3} \left\langle \Delta_{\alpha\beta}k^{\alpha }k^{\beta
} \mathcal{S}_{\mathbf{k}}^{\left( 0 \right) }  \right\rangle_{0}, \\
\kappa_n &=& \frac{1}{3} \left\langle \Delta_{\alpha\beta}k^{\alpha }k^{\beta} \mathcal{S}_{\mathbf{k}}^{\left( 1 \right) }  \right\rangle_{0}, \\
\eta &=& \frac{1}{15} \left\langle (\Delta_{\alpha\beta}k^{\alpha }k^{\beta})^2 \mathcal{S}_{\mathbf{k}}^{\left( 2 \right) }   \right\rangle_{0}.   \label{definitions}
\end{eqnarray}
The shear viscosity depends only on $\mathcal{S}_{\mathbf{k}}^{\left( 2 \right) }$ and, thus, it will not be modified by our new relaxation time approximation. Given that most previous works employing the relaxation time approximation focused on the effects of shear viscosity, the fundamental problems in the AW approximation discussed here were not evident. On the other hand, the bulk viscosity and particle diffusion coefficients depend on $\mathcal{S}_{\mathbf{k}}^{\left( 0 \right) }$ and $\mathcal{S}_{\mathbf{k}}^{\left( 1 \right) }$ and, thus, they are sensitive to the inconsistencies of the AW approximation. As a matter of fact, we shall demonstrate that the effects of our corrections are significant for those coefficients.

\begin{figure}[!h]
    \centering
\begin{subfigure}{0.47\textwidth}
  \includegraphics[width=\linewidth]{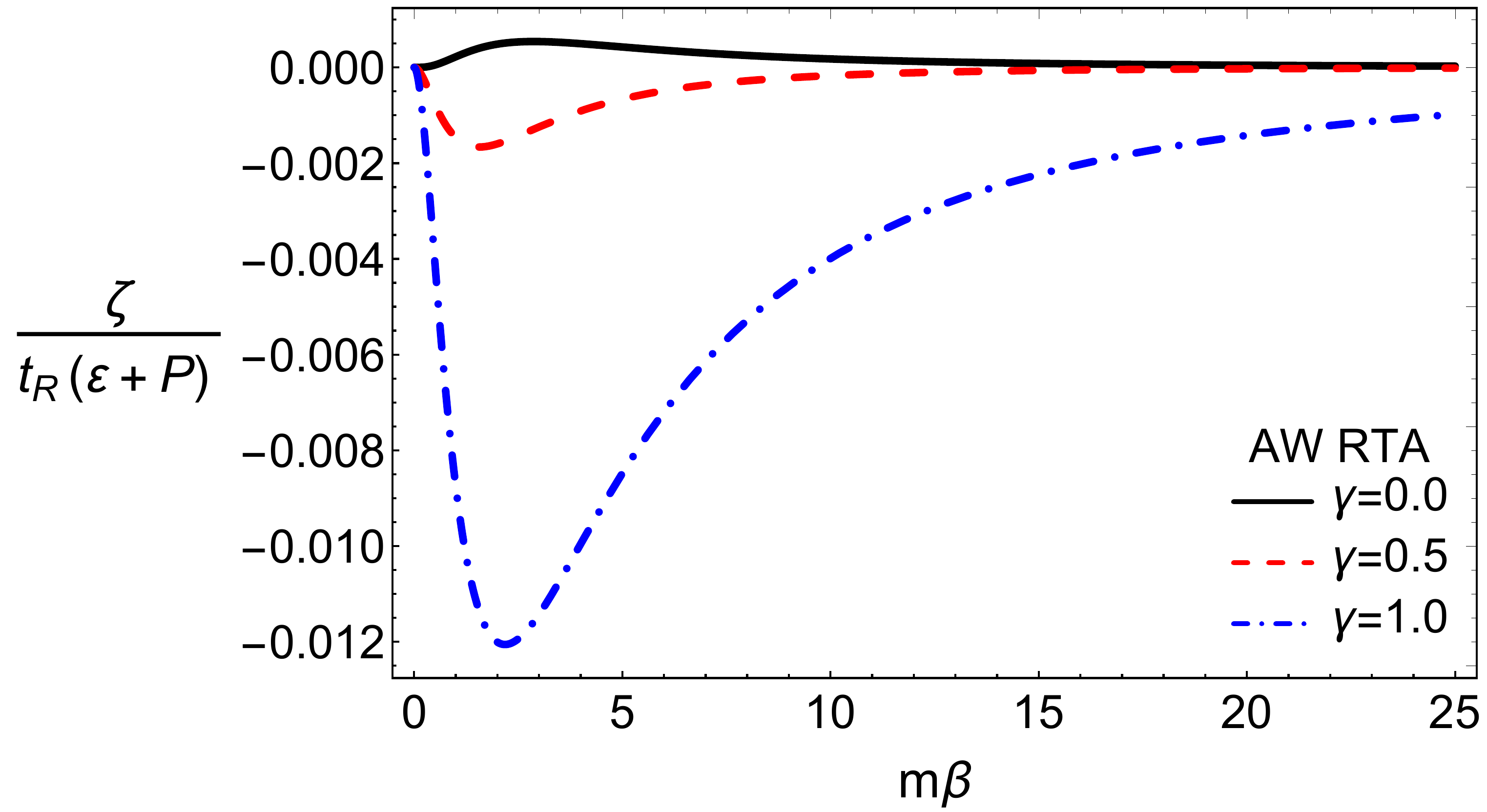}
\end{subfigure}\hfil 
\begin{subfigure}{0.47\textwidth}
  \includegraphics[width=\linewidth]{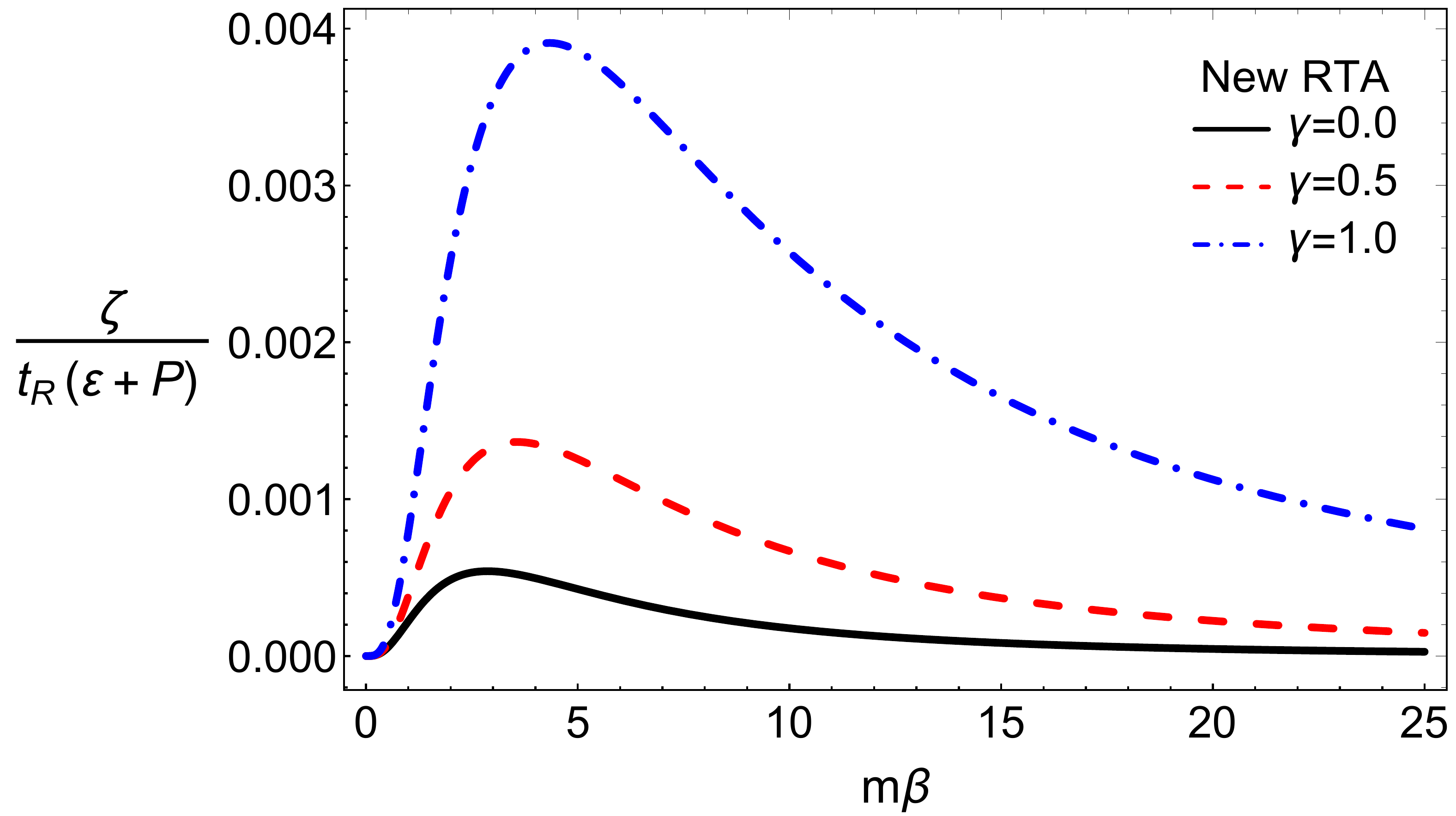}
\end{subfigure}\hfil 

\caption{Dimensionless bulk viscosity coefficient, $\zeta/[t_{R} (\varepsilon + P)]$, as a function of $m \beta$, assuming Landau matching conditions. The results are shown for $\gamma=0$ (solid curve), $\gamma=0.5$ (dashed curve), and $\gamma=1$ (dashed-dotted curve). }
\label{fig:1}
\end{figure}

\begin{figure}[!h]
    \centering
\begin{subfigure}{0.42\textwidth}
  \includegraphics[width=\linewidth]{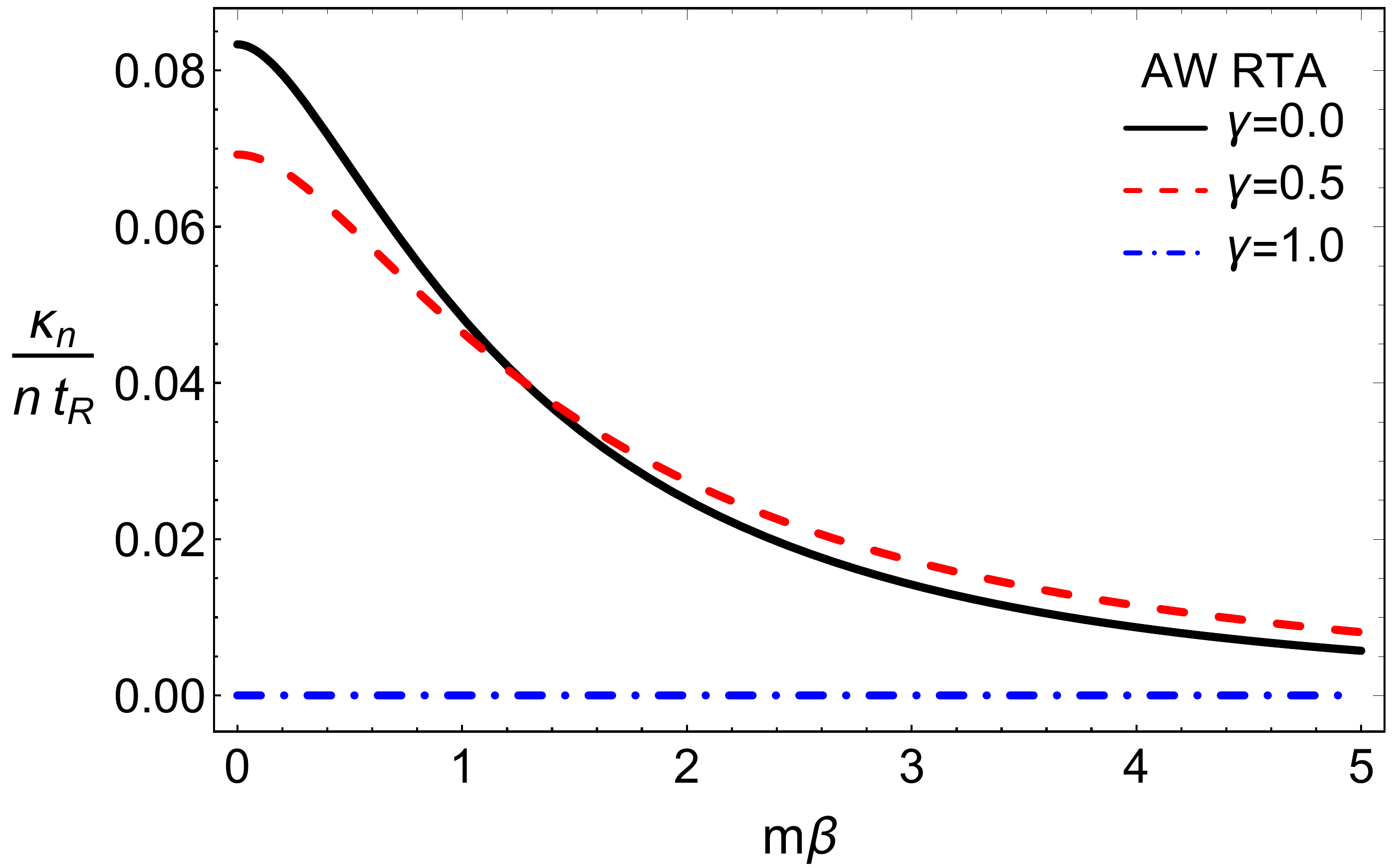}
\end{subfigure}\hfil 
\begin{subfigure}{0.42\textwidth}
  \includegraphics[width=\linewidth]{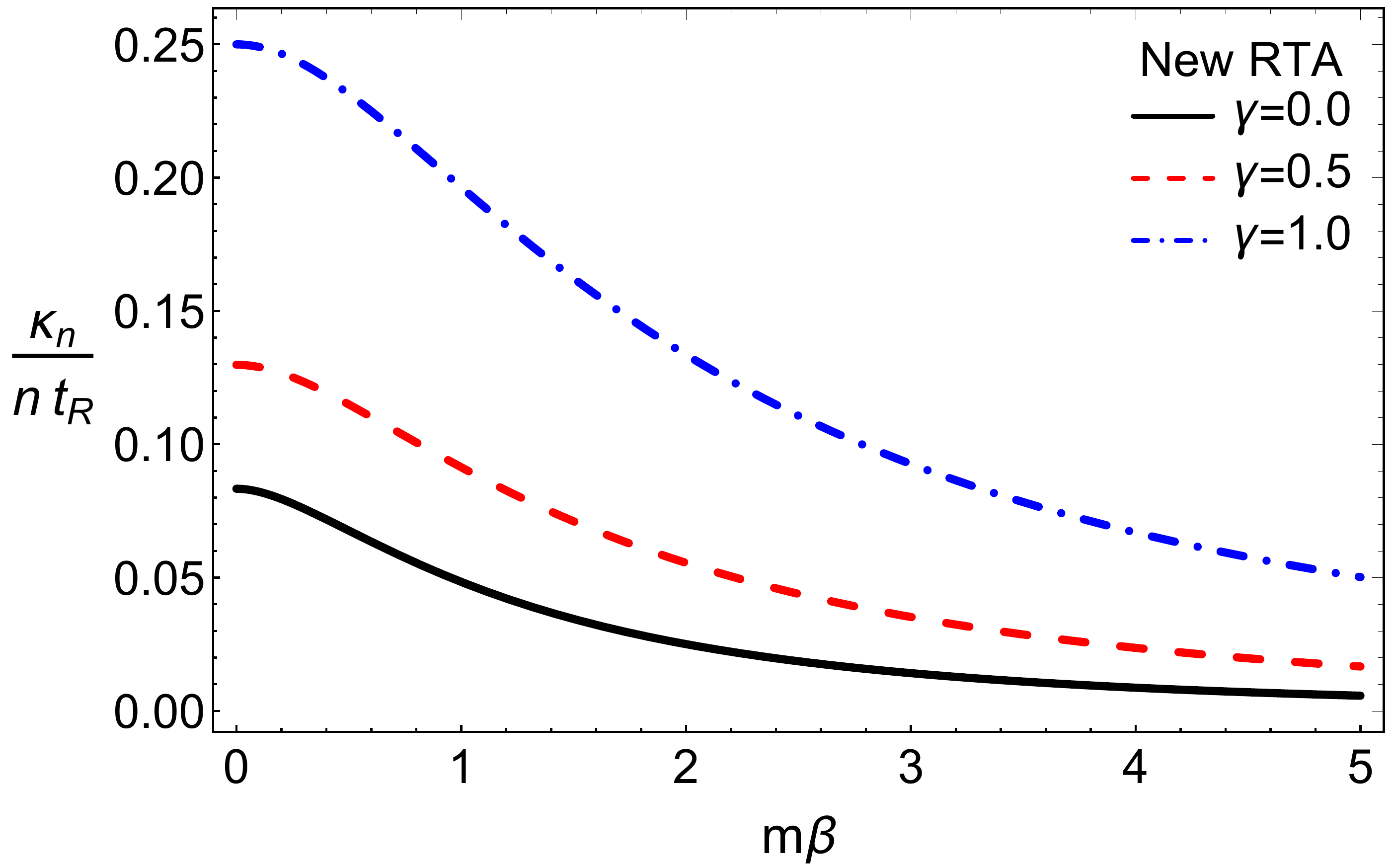}
\end{subfigure}\hfil 
\caption{Dimensionless particle diffusion coefficient, $\kappa_n/(t_{R} n)$, as a function of $m \beta$, assuming Landau matching conditions. The results are shown for $\gamma=0$ (solid curve), $\gamma=0.5$ (dashed curve), and $\gamma=1$ (dashed-dotted curve). }
\label{fig:2}
\end{figure}

In the following, we calculate $\zeta$ and $\kappa_n$ in the relaxation time approximation. For this purpose we assume Landau matching conditions (which corresponds to setting $g_{\mathbf{k}}=q_{\mathbf{k}}=E_{\mathbf{k}}$ and $h_{\mathbf{k}}=E_{\mathbf{k}}^2$) and, motivated by Refs.\ \cite{Dusling:2009df,Dusling:2011fd,Kurkela:2017xis}, we take the following momentum dependence for the relaxation time,
\begin{equation}
\tau_R = (\beta E_{\mathbf{k}})^\gamma t_R,
\end{equation}
where the coefficient $t_R$ carries no momentum dependence and $\gamma$ is an arbitrary constant that will control the energy dependence of the relaxation time. For the sake of illustration, we shall consider results for $\gamma=0$ (which is equivalent to the AW approximation, when assuming Landau matching conditions), $\gamma=0.5$, and $\gamma=1$. Results for the bulk viscosity are shown in Fig.~\ref{fig:1} for the traditional RTA (left panel) and our novel RTA (right panel), while results for the particle diffusion are shown in Fig.~\ref{fig:2} for the traditional RTA (left panel) and our novel RTA (right panel).

We see that the modifications we implemented to the relaxation time approximation affect considerably the bulk and particle diffusion coefficients, when an energy dependent relaxation time is employed. In fact, as shown in Fig.\ \ref{fig:1}, the traditional RTA can lead to negative transport coefficients depending on the choice of $\gamma$  -- a clear violation of the second law of thermodynamics. For example, if we take $\gamma=0.5$, a value argued to be a good approximation for effective kinetic descriptions of
quantum chromodynamics \cite{Dusling:2009df,Dusling:2011fd,Kurkela:2017xis}, the bulk viscosity becomes negative. Furthermore, we note that the matching conditions will also affect these transport coefficients. However, we leave such study to future work.\\

\noindent
\textit{Conclusions:} We showed that the widely employed relaxation time approximation to the relativistic Boltzmann equation, derived in \cite{AW}, is inconsistent with microscopic and macroscopic conservation laws. We proposed a novel relaxation time approximation that is free from such flaws and can be applied to describe systems with arbitrary relaxation times and matching conditions. We further demonstrated that the modifications we proposed considerably affected the temperature dependence of the bulk viscosity and particle diffusion coefficients. 

The relaxation time formalism presented here is the only one that can be applied to investigate how different choices of matching conditions (i.e., different hydrodynamic frames) affect the physical and mathematical properties of relativistic hydrodynamics. This will be instrumental when investigating causality and stability of relativistic fluids defined via a generalized derivative expansion, as shown in \cite{Bemfica:2017wps,Kovtun:2019hdm,Bemfica:2019knx,Hoult:2020eho,Bemfica:2020zjp}. Finally, it is essential to investigate how the analytic and semi-analytic solutions of rapidly expanding gases, derived in Refs.~\cite{Denicol:2014xca,Denicol:2014tha,Noronha:2015jia,Hatta:2015kia,Florkowski:2013lya,Bazow:2015dha,Bazow:2016oky}, are modified in our novel relaxation time approximation, in particular when other matching conditions are employed. This will be important in understanding the emergence of hydrodynamic behavior in ultrarelativistic heavy ion collisions. \\

\noindent
\textit{Acknowledgements:} G.~S.~R.~and G.~S.~D.~acknowledge Conselho Nacional de Desenvolvimento Cient\'ifico e Tecnol\'ogico (CNPq) and Funda\c c\~ao Carlos Chagas Filho de Amparo \`a Pesquisa do Estado do Rio de Janeiro (FAPERJ) for financial support. J.~N.~is partially supported by the U.S.~Department of Energy, Office of Science, Office for Nuclear Physics under Award No. DE-SC0021301.

\end{document}